\documentclass[twocolumn]{aastex631}

\usepackage{xcolor}

\usepackage{threeparttable}
\usepackage{pifont}
\usepackage{float}
\usepackage{soul}

\newcommand{\red}[1]{#1}

\def\arc{^{\prime\prime}}
\def\zphot{\ifmmode z_{\rm phot}\else$z_{\rm phot}$\fi}

\def\Hi{H\,{\sc i}}
\def\Hii{H\,{\sc ii}}
\def\Oi{[O\,{\sc i}]}
\def\Oium{[O\,{\sc i}]\,$145\,{\rm \mu m}$}

\def\Oiii{[O\,{\sc iii}]}
\def\Cii{[C\,{\sc ii}]}
\def\Ciium{[C\,{\sc ii}]\,$158\,{\rm \mu m}$}
\def\Oi{[O\,{\sc i}]}
\def\Oiiium{[O\,{\sc iii}]\,$88\,{\rm \mu m}$}
\def\Nii{[N\,{\sc ii}]}
\def\Niium{[N\,{\sc ii}]\,$205\,{\rm \mu m}$}

\def\ltsima{$\buildrel<\over\sim$}
\def\la{\lower.5ex\hbox{\ltsima}~}
\def\gtsima{$\buildrel>\over\sim$}
\def\ga{\lower.5ex\hbox{\gtsima}~}
\def\deg~{$^{\circ}$}

\def\aap{A\&A}

\begin{document}

\title{ALMA Observations of \Oium\ and \Niium\ Emission lines from Star-Forming Galaxies at $z\sim7$}

\correspondingauthor{Yoshinobu Fudamoto}
\email{y.fudamoto@chiba-u.jp}

\author[0000-0001-7440-8832]{Yoshinobu Fudamoto} 
\affiliation{Center for Frontier Science, Chiba University, 1-33 Yayoi-cho, Inage-ku, Chiba 263-8522, Japan}
\affiliation{Waseda Research Institute for Science and Engineering, Faculty of Science and Engineering, Waseda University, 3-4-1 Okubo, Shinjuku, Tokyo 169-8555, Japan}
\affiliation{National Astronomical Observatory of Japan, 2-21-1, Osawa, Mitaka, Tokyo, Japan}

\author[0000-0002-7779-8677]{Akio K. Inoue} 
\affiliation{Waseda Research Institute for Science and Engineering, Faculty of Science and Engineering, Waseda University, 3-4-1 Okubo, Shinjuku, Tokyo 169-8555, Japan}
\affiliation{Department of Physics, School of Advanced Science and Engineering, Faculty of Science and Engineering, Waseda University, 3-4-1, Okubo, Shinjuku, Tokyo 169-8555, Japan}

\author[0000-0002-4989-2471]{Rychard Bouwens}
\affiliation{Leiden Observatory, Leiden University, NL-2300 RA Leiden, Netherlands}

\author[0000-0003-4268-0393]{Hanae Inami}
\affiliation{Hiroshima Astrophysical Science Center, Hiroshima University, 1-3-1 Kagamiyama, Higashi-Hiroshima, Hiroshima 739-8526, Japan}

\author[0000-0001-8034-7802]{Renske Smit}
\affiliation{Astrophysics Research Institute, Liverpool John Moores University, 146 Brownlow Hill, Liverpool L3 5RF, United Kingdom}

\author{Dan Stark}
\affiliation{Steward Observatory, University of Arizona, 933 N Cherry
Ave, Tucson, AZ 85721, United States}

\author[0000-0002-6290-3198]{Manuel Aravena}
\affiliation{Instituto de Estudios Astrof\'{\i}cos, Facultad de Ingenier\'{\i}a y Ciencias, Universidad Diego Portales, Av. Ej\'ercito 441, Santiago, Chile}
\affiliation{Millenium Nucleus for Galaxies (MINGAL)}

\author[0000-0002-7129-5761]{Andrea Pallottini}
\affiliation{Scuola Normale Superiore, Piazza dei Cavalieri 7, 56126 Pisa, Italy}

\author[0000-0002-0898-4038]{Takuya Hashimoto}
\affiliation{Division of Physics, Faculty of Pure and Applied Sciences, University of Tsukuba, Tsukuba, Ibaraki 305-8571, Japan}
\affiliation{Tomonaga Center for the History of the Universe (TCHoU), Faculty of Pure and Applied Sciences, University of Tsukuba, Tsukuba, Ibaraki 305-8571, Japan}

\author[0000-0003-3484-399X]{Masamune Oguri}
\affiliation{Center for Frontier Science, Chiba University, 1-33 Yayoi-cho, Inage-ku, Chiba 263-8522, Japan}

\author[0000-0002-4205-9567]{Hiddo Algera}
\affiliation{Institute of Astronomy and Astrophysics, Academia Sinica, 11F of Astronomy-Mathematics Building, No.1, Sec. 4, Roosevelt Rd, Taipei 106216, Taiwan, R.O.C.}

\author[0000-0003-3917-1678]{Rebecca A. A. Bowler}
\affiliation{Jodrell Bank Centre for Astrophysics, Department of Physics and Astronomy, School of Natural Sciences, The University of Manchester, Manchester, M13 9PL, UK}

\author[0000-0001-9759-4797]{Elisabete da Cunha}
\affiliation{International Centre for Radio Astronomy Research, University of Western Australia, 35 Stirling Hwy, Crawley, WA 6009, Australia}
\affiliation{ARC Centre of Excellence for All Sky Astrophysics in 3 Dimensions (ASTRO 3D), Australia}

\author[0000-0001-8460-1564]{Pratika Dayal}
\affiliation{Kapteyn Astronomical Institute, University of Groningen, P.O. Box 800, 9700 AV Groningen, The Netherlands}

\author[0000-0002-9400-7312]{Andrea Ferrara}
\affiliation{Scuola Normale Superiore, Piazza dei Cavalieri 7, 56126 Pisa, Italy}

\author[0000-0001-7201-5066]{Seiji Fujimoto}
\affiliation{Department of Astronomy, The University of Texas at Austin, Austin, TX 78712, USA}

\author[0000-0002-9389-7413]{Kasper E. Heintz}
\affil{Cosmic Dawn Center (DAWN), Denmark}
\affil{Niels Bohr Institute, University of Copenhagen, Jagtvej 128, 2200 Copenhagen N, Denmark}

\author[0000-0002-6488-471X]{Alexander P. S. Hygate}
\affiliation{Leiden Observatory, Leiden University, NL-2300 RA Leiden, Netherlands}

\author[0009-0005-6803-6805]{Ivana F. van Leeuwen}
\affiliation{Leiden Observatory, Leiden University, NL-2300 RA Leiden, Netherlands}

\author[0000-0001-9419-6355]{Ilse De Looze}
\affiliation{Sterrenkundig Observatorium, Ghent University, Krijgslaan 281 - S9, B-9000 Gent, Belgium}

\author[0009-0009-2671-4160]{Lucie E. Rowland}
\affiliation{Leiden Observatory, Leiden University, P.O. Box 9513, 2300 RA Leiden, The Netherlands}

\author[0000-0001-7768-5309]{Mauro Stefanon}
\affiliation{Departament d'Astronomia i Astrof{\'i}sica, Universitat de Val{\`e}ncia, C. Dr. Moliner 50, E-46100 Burjassot, Val{\`e}ncia, Spain}
\affiliation{Unidad Asociada CSIC ``Grupo de Astrofísica Extragal{\'a}ctica y Cosmolog{\'i}a'' (Instituto de F{\'i}sica de Cantabria - Universitat de Val{\`e}ncia)}

\author[0000-0001-6958-7856]{Yuma Sugahara}
\affiliation{Waseda Research Institute for Science and Engineering, Faculty of Science and Engineering, Waseda University, 3-4-1 Okubo, Shinjuku, Tokyo 169-8555, Japan}
\affiliation{National Astronomical Observatory of Japan, 2-21-1, Osawa, Mitaka, Tokyo, Japan}

\author[0000-0002-7595-121X]{Joris Witstok}
\affiliation{Kavli Institute for Cosmology, University of Cambridge, Madingley Road, Cambridge, CB3 0HA, UK}
\affiliation{Cavendish Laboratory, University of Cambridge, 19 JJ Thomson Avenue, Cambridge, CB3 0HE, UK}

\author[0000-0001-5434-5942]{Paul P. van der Werf}
\affiliation{Leiden Observatory, Leiden University, P.O. Box 9513, 2300 RA Leiden, The Netherlands}



\begin{abstract}
We present results of new observations of \Oium\ and \Niium\ emission lines from four star-forming galaxies at redshifts between $z=6.58$ and $7.68$ that have previous detections of \Ciium\ and dust continua.
Using ALMA, we successfully detect \Oium\ emission from all targets at $>4\,\sigma$ significance.
However, \Niium\ emission is undetected in all galaxies (SNR $<3.5\,\sigma$) except for a tentative detection from A1689-zD1.
From the observed high \Cii/\Nii\ emission line ratios ($\gtrsim20 - 80$), we find that most of the \Ciium\ emission arise from neutral gas regions ($3\,\sigma$ lower limits of $\gtrsim 74 - 96\%$).
From \Oium, \Ciium\ lines, and infrared luminosities, we estimate the neutral gas densities of $n_{\rm H}=10^{3.5}$ - $10^6\,{\rm cm^{-3}}$ and the far-ultraviolet (FUV) radiation strengths of $G_0\sim10^{2.5}$-$10^{3}$.
While the neutral gas densities are similar to those of high-redshift starburst galaxies, the FUV strengths are lower compared to both local and high-redshift starbursts. 
Finally, we estimate atomic hydrogen masses using \Oium\ emission lines and the oxygen abundances measured from recent JWST observations. We find gas mass ratios of $f_{\rm gas}\sim0.3$ - $0.8$, which are similar to earlier studies using \Ciium.
Starting from this pilot observation, future large \Oium\ emission line surveys will provide us with currently little-known neutral gas properties of star-forming galaxies in the early Universe.
\end{abstract}



\section{Introduction} \label{sec:intro}
Studying how star-forming galaxies evolved in the epoch of reionization (EoR; e.g., at $z>6$) has crucial implications for our understanding of the cosmic reionization \citep[e.g.,][]{Dayal2018,Robertson2022} as well as for our understanding of the galaxy evolution in the later epochs \citep[e.g.,][]{Madau2014}.
Deep observations using the Hubble Space Telescope, the recent observations using the James Webb Space Telescope (JWST), and several ground-based optical/near-infrared (NIR) facilities have completely revolutionized our view of galaxy evolution in the high-redshift Universe through their extremely high sensitivity and high resolution in the rest-frame ultraviolet (UV) and/or optical wavelengths \citep[e.g.,][]{Bouwens2015,Oesch2018,Naidu2022,Oesch2023,Cameron2023,Finkelstein2023,Fujimoto2023,Harikane2023,Robertson2023,Rieke2023,Williams2023,Wang2024,Carniani2024}.
While these rest-frame UV/optical observations enabled access to the galaxies' star-formation activities through the observations of ionized gas, the cold neutral and the molecular gas -- the direct fuel for star formation activities -- is challenging to probe in emission at the rest-frame UV/optical wavelengths. This is because cooling lines from the neutral gas emits most of its energy at far-infrared (FIR) wavelength range (however, see e.g., \citealt[][]{Heintz2023} for probing the neutral gas in absorption).

The unprecedented sensitivity and high spatial resolution of the Atacama Large Millimeter/submillimeter Array (ALMA) has largely improved studies of galaxy formation and evolution by enabling to probe dust continuum and emission lines in the far-infrared (FIR) wavelength from galaxies in the EoR (see \citealt{Hodge2020} for a review).
In particular, recent ALMA observations detected \Cii\,$158\,{\rm \mu m}$ emission lines from a statistically significant number of EoR galaxies \citep[e.g.,][]{Smit2018,Hashimoto2019,Carniani2020,Bouwens2022,Wong2022,Schouws2023,Fujimoto2022,Heintz2023b,2024A&A...682A.166F}.
Observations of the \Cii\ emission line and studies of the interstellar medium (ISM) by combining multiple emission lines provide crucial diagnostics for the physical properties and galaxies' star formation rates \citep[e.g.,][]{Harikane2020,Schaerer2020,Sugahara2022,Schouws2022,Witstok2022,Algera2023,Killi2023,Vallini2021,Vallini2024}.
However, multiple FIR emission line observations of high-redshift galaxies are limited to extremely IR luminous galaxies such as submillimeter galaxies \citep[SMGs: e.g.,][]{DeBreuck2019,Meyer2022} and the numerous more representative star-forming galaxies were typically not targeted at high redshift.
To change this situation, in this paper, we study the ISM properties using multiple FIR emission lines of typical star-forming galaxies in the EoR.

\red{Firstly, we examine the fraction of \Ciium\ originating from either ionized gas or neutral gas (both atomic and molecular) using observations of the \Niium\ emission line.
C$^+$ ions exist in both ionized and neutral gas because their ionization potential is $11.3\,{\rm eV}$, while N$^+$ is found only in ionized gas due to its  higher ionization potential of $14.5\,{\rm eV}$, which exceeds that of hydrogen.
Because \Niium\ and \Ciium\ have similar critical densities ($n_{\rm crit}\sim40\,{\rm cm^{-3}}$ and $n_{\rm crit}\sim50\,{\rm cm^{-3}}$ with electrons, respectively) and excitation temperatures ($\sim70\,{\rm K}$ for \Niium\ and $\sim50\,{\rm K}$ for \Ciium), both lines are produced with comparable luminosities inside ionized gas clouds where ionized nitrogen is present.
Thus, the line ratio between \Ciium\ and \Niium\ for ionized region can be estimated based on their assumed abundance and emissivity and the excess \Ciium\ luminosity would originate from neutral gas.}

As several previous studies performed for local and high-redshift IR luminous galaxies, the luminosity ratio between \Niium\ and \Ciium\ indeed works as an indicator for constraining the origin of \Ciium\ emission line for low- and high-redshift star-forming galaxies \citep{Malhotra2001,Oberst2011,Decarli2014,Bethermin2016,Pavesi2016,Croxall2017, Lee2021,Witstok2022,Pensabene2021,Decarli2023}.

Secondly, we study the neutral gas properties of star-forming galaxies in the EoR using the \Oium\ observations.
Having similar ionization potential as that of hydrogen ($13.62\,{\rm eV}$ for Oxygen), neutral oxygen, O$^0$, is an ideal probe for the neutral star-forming gas.
The \Oi\,$145\,{\rm \mu m}$ emission line ($\lambda_{\rm rest}= 145.53\,{\rm \mu m}$) is one of the \red{important} coolants in neutral gas, trace up to dense gas as it has a high critical densities of ${\rm n^{H}_{crit}} \sim 10^5\,{\rm cm^{-3}}$, and is optically thin unlike \Oi$\,63\,{\rm \mu m}$ \citep[e.g.,][]{Tielens1985,Kaufman1999}.
\setstcolor{magenta}
\red{The \Oium\ line has been observed in the local Universe} \citep[e.g.,][]{Diaz-Santos2017,Herrera-Camus2018} as well as from high-redshift starbursts and active galactic nuclei \citep[][]{DeBreuck2019,Yang2019,Novak2020,Lee2021,Pensabene2021,Meyer2022,Litke2023}. These studies showed that the combination of \Ciium\ and \Oium\ emission lines is sensitive to the neutral gas density ($n$), and far-UV radiation field ($G_0$), enabling a measurement of these essential ISM parameters.
 \red{While several studies have highlighted the importance of \Oium\ emission lines in high-redshift SMGs and QSO hosts \citep[e.g.,][]{DeBreuck2019,Lee2021,Meyer2022}, no observations of \Oi\ have been performed for star-forming galaxies that are more representative of the typical galaxy population in the EoR.}
In this study, we present the analyses and results of the first pilot observations for the \Oium\ emission lines in \red{\Cii\ bright} star-forming galaxies in the EoR.

This paper is organized as follows: in \S\ref{sec:obs} we describe our target galaxies and the ALMA observations used in this study. In \S\ref{sec:analysis}, we present our data analysis and measurements for the \Oium\ and \Niium\ emission lines.  In \S\ref{sec:results}, we present the results of the study. In \S\ref{sec:discussion}, we compare to previous studies and discuss our results. Finally, we conclude with the summary in \S\ref{sec:conclusion}.
Throughout this paper, we assume a cosmology with $(\Omega_m,\Omega_{\Lambda},h)=(0.3,0.7,0.7)$, and the \citealt{Chabrier2003} initial mass function (IMF) with stellar masses ranging from $0.1$ to $300\,{\rm M_{\odot}}$, where applicable.

\section{Observation} \label{sec:obs}

\renewcommand{\arraystretch}{1.2}
\begin{table*}
    \centering
    \caption{List of the target galaxies in this study, ordered by their redshifts. Properties for REBELS-38, REBELS-25, REBELS-18 are adopted from \citealt{Bouwens2022} and Schouws et al. in preparation, and A1689-zD1 are presented in \citealt{Bradley2008,Watson2015,Inoue2020,Bakx2021}. Magnification of A1689-zD1 is newly updated in this study (see Appendix \ref{sec:lens}). }
    \label{tab:targets}
    \begin{tabular}{lccccccc}
         \hline
         Source &  $\alpha_{\rm J2000}$ (deg) & $\delta_{\rm J2000}$ (deg) & $z_{\rm spec}$ & Magnification ($\mu$) & $L_{\rm UV}\,(10^{11}\,{\rm L_{\odot}})$ & $M_{\ast}\,(10^{9}\,{\rm M_{\odot}})$ & SFR (${\rm M_{\odot}\,yr^{-1}}$)\\
         \hline
         REBELS-38 & 150.725208 & 2.703333 & 6.577 & -- & $1.2\pm0.3$ & $3.8^{+1.7}_{-3.6}$ & $116\pm45$ \\
         A1689-zD1 & 197.874708 & -1.321861 & 7.132 & \red{$4.14\pm0.36$} &  \red{$0.63\pm0.05$$^{\dagger}$} & \red{$3.7^{+1.6}_{-1.1}$$^{\dagger}$} & \red{$83\pm8^{\dagger}$} \\
         REBELS-25 & 150.134667 & 1.742028 & 7.307 & -- & $1.0\pm0.2$ & $7.7^{+3.2}_{-2.6}$ & $200\pm83$\\
         REBELS-18 & 149.449583 & 2.345472 & 7.675 & -- & $1.5\pm0.2$ & $3.1^{+8.1}_{-2.5}$ & $72\pm20$\\
        \hline
        \multicolumn{7}{l}{$\dagger$ Values are corrected by scaling with the updated gravitational lensing magnification factor of \red{$\mu=4.14$}.}\\
        \multicolumn{7}{l}{\red{(see Appendix \ref{sec:lens})}}
    \end{tabular}
\end{table*}

\begin{table*}
    \centering
    \caption{Summary of ALMA observations presented in this work}
    \label{tab:ALMAobs}
    \begin{tabular}{ccccccccc}
        \hline
        Source & Emission Line & \red{$\nu_{\rm obs}^{\ast}$} & $t_{\rm int}$ & PVW & Beam \red{FWHM} & RMS$_{\rm cont.}$ & RMS$_{\rm line, 25km/s}$  & ALMA project ID\\
         & & (GHz) & (hrs) & $\rm{(mm)}$ & & ($\rm{\mu Jy/beam}$) & ($\rm{mJy/beam}$) & \\
        \hline
            REBELS-38 & \Niium\ & $192.8$ & $1.5$ & 0.749 & $1\farcs2\times0\farcs9$ & $11$ & $0.17$ & \#2022.1.01384.S\\
            & \Oium\ & $271.9$ & $1.2$ &  $0.322$ & $1\farcs3\times1\farcs0$ & $11$ & $0.24$ & \#2022.1.00446.S\\
            A1689-zD1 & \Niium$^{\ast\ast}$ & $179.7$ & $2.3$ & $0.503$ & $1\farcs5\times1\farcs2$ & $6.5$ & $0.10$ & \#2021.1.00247.S\\
            & \Oium\ & $252.3$ & $0.2$ & $0.587$ & $1\farcs3\times1\farcs0$ & $25$ & $0.44$ & \#2022.1.00446.S\\
            REBELS-25 & \Niium & $175.9$ & $1.5$ & 0.682 & $1\farcs5\times1\farcs0$ & $8.8$ & $0.28$ & \#2022.1.01384.S\\
            & \Oium\ & $248.0$ & $0.8$ & $1.470$ & $0\farcs6\times0\farcs5$ & $14$ & $0.27$ & \#2022.1.00999.S\\
            REBELS-18 & \Oium\ & $237.5$ & $2.6$ & $0.352$ & $1\farcs3\times1\farcs2$ & $6.5$ & $0.12$ & \#2022.1.00446.S\\
        \hline
        \multicolumn{9}{l}{\red{$\ast$ Target frequency of observations. See Table \ref{tab:ALMAmeasurements} for measured redshifts of each emission line.}}\\
        \multicolumn{9}{l}{$\ast\ast$ \Niium\ observation of A1689-zD1 is further presented and discussed in detail in Fujimoto et al. in preparation.}\\
    \end{tabular}
\end{table*}

\subsection{Target Galaxies}
Table \ref{tab:targets} summarizes our target galaxies and their basic properties.
Below, we details our sample selection criteria.

Our targets are star-forming galaxies at $z>6.5$, i.e., those that have star-formation rates and stellar masses consistent with the main-sequence of star-forming galaxies at $z\sim7$ \citep{Topping2022}.
To select targets, we searched all previous \Cii\ emission line observations including the cycle-7 ALMA large program Reionization Era Bright Emission Line Survey (REBELS; \citealt{Bouwens2022}) as well as previously published \Cii\ emission line observations \citep[e.g.,][]{Smit2018, Hashimoto2019, Carniani2020,Akins2022,Schouws2022,Schouws2025}.
Individual studies have their own selection methods based on the estimated accuracy of photometric redshifts as well as based on detections of Ly$\alpha$ emission lines. However, these galaxies were originally all selected based on their bright rest-frame UV luminosities (typically $M_{\rm UV} < -21\,{\rm mag}$; \citealt{Bouwens2022}).

We further selected galaxies based on their expected luminosities of \Oium\ and \Niium\ emission based on previously detected \Ciium\ line emission luminosities.
In particular, we performed flux limited selection of galaxies with their apparent \Cii\ luminosity of $L_{\rm [CII]}>1.1\times10^{9}\,{\rm L_{\odot}}$.
Using these criteria, we selected four \Cii\ luminous star-forming galaxies; REBELS-38 at $z=6.577$ \citep{Bouwens2022,Algera2023}, A1689-zD1 at $z=7.132$ \citep[e.g.,][]{Watson2015,Bakx2021,Akins2022,Killi2023}, REBELS-25 at $z=7.307$ \citep{Bouwens2022,Hygate2023,Algera2023,Rowland2024}, and REBELS-18 at $z=7.675$ \citep{Bouwens2022}.

\subsection{ALMA Observations}
Observation conditions, integration times, and beam sizes of these observations are summarized in Table \ref{tab:ALMAobs}. In the following, we present details of these observations.

\subsubsection{\Oium\ and \Niium\ Emission Line}
Observations of \Oium\ emission lines of target galaxies were newly performed during ALMA Cycle-9 \#2022.1.00446.S (PI: Fudamoto) and \#2022.1.00999.S (PI: De Looze). \Niium\ emission lines were newly observed in Cycle-9 \#2022.1.01384.S (PI: Fudamoto) for REBELS-38 and REBELS-25. \Niium\ data of A1689-zD1 were retrieved from the ALMA archive \#2021.1.00247.S (PI: Fujimoto). 
\Niium\ emission line observations of REBELS-38 and A1689-zD1 were fully executed. On the other hand, \Niium\ observations of REBELS-25 were only partially performed.

\subsection{ALMA data reduction}

\red{In this study, we present newly obtained ALMA data for the \Oi\ and \Nii\ emission lines from the target galaxies.}
\Ciium\ and \Oiiium\ emission line data are adopted from \citet{Watson2015,Akins2022, Bouwens2022,Killi2023,Wang2024,Algera2024}.

The observed ALMA data of \Oium\ and \Niium\ were reduced and analyzed using the Common Astronomy Software Application (CASA) version 6.2.1 \citep{Bean2022}.
ALMA observations for \Oium\ and \Niium\ emission lines were calibrated using the second stage quality assurance (QA2) pipeline codes (i.e., \path{ScriptForPI.py}) without further flagging and editing.

To examine and analyze emission lines, we created data cubes using spectral windows (SPWs) that include \red{the sky frequencies of each emission line, accounting for the target's spectroscopic redshift.}
To create data cubes using  the \path{tclean} task of CASA, we applied a fixed velocity channel binning of $25\,{\rm km/s}$ for all data cubes.
Rest-frame frequencies of each data cube were set to the sky frequency of each emission line observation based on the \Cii\ emission line detections (table \ref{tab:ALMAmeasurements}).
Underlying dust continua of the target galaxies were subtracted using CASA task \path{uvcontsub} by fitting flat continua for the SPWs that do not include emission lines.
\red{Dust continuum maps of all observations were made using the continuum SPWs and using the \path{tclean} task of CASA and using the multi-frequency synthesis (\path{mfs}) imaging mode.}

When we created data cubes and dust continuum maps, we performed synthesized beam deconvolutions using the \path{tclean} by setting stopping thresholds to be $3\times$ the pixel-by-pixel root-mean-square (RMS) of the dirty images and by setting the maximum number of iterations to 500 times.
For data cubes, the background RMS was calculated as averages of RMS of all channels.
The threshold and parameters did not affect final data quality as there were no particularly bright sources in the fields of views.
All data cubes and continuum maps were imaged using the \path{natural} weighting scheme to maximize point source sensitivity.
The achieved sensitivity of pixel-by-pixel RMS of the cleaned cubes and continuum maps are presented in table \ref{tab:ALMAobs}.

\section{Analysis} \label{sec:analysis}

\begin{figure*}[tb]
    \centering
    \includegraphics[width=0.95\textwidth]{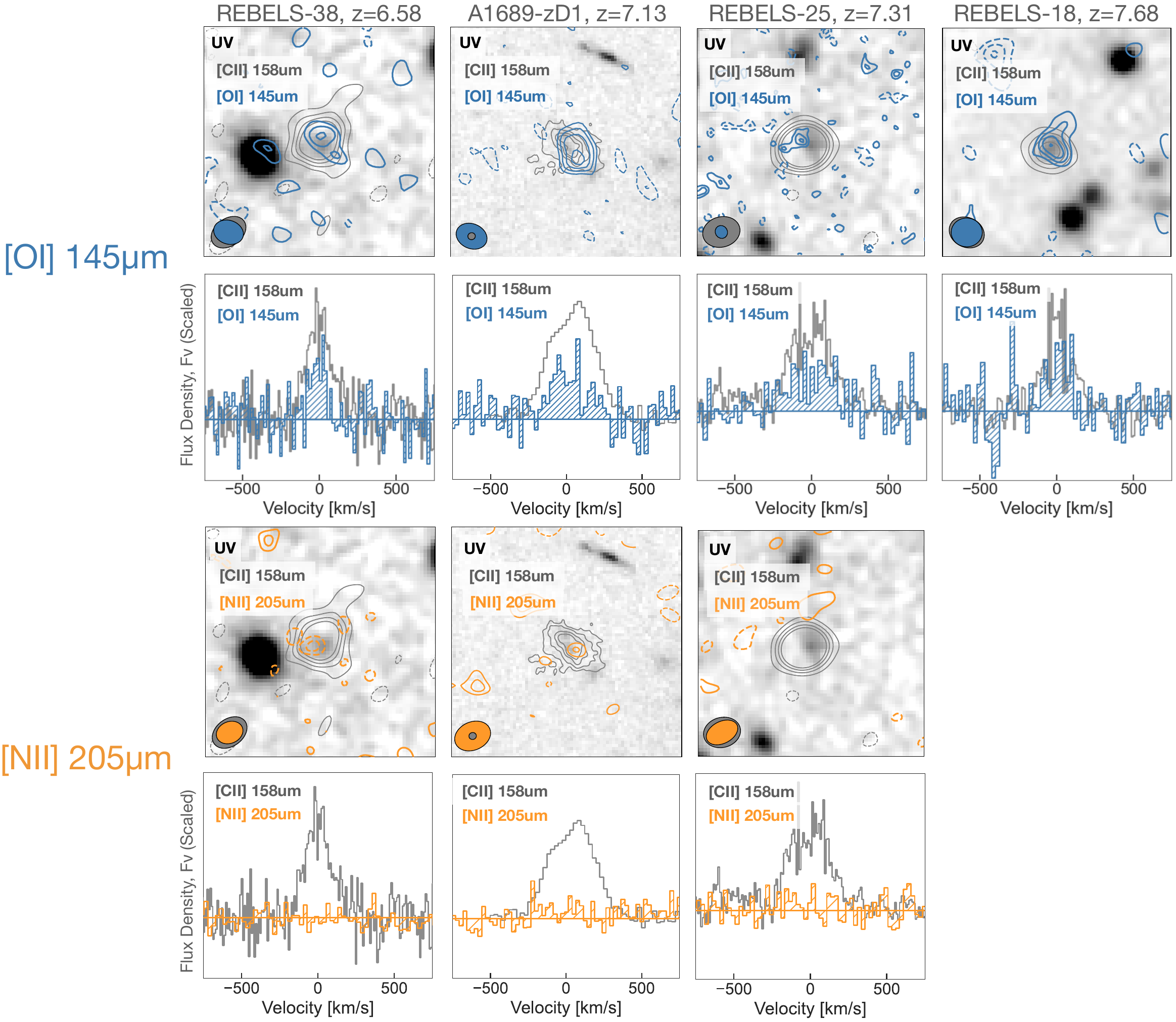}
    \caption{Cutouts and spectra of \Ciium, \Oium, and \Niium\ emission lines of the targets.  For REBELS-38, REBELS25, and REBELS-18, background images of cutouts show rest-frame UV continuum obtained by stacking J-, H-, and K-band images from UVISTA survey \citep{McCracken2012}. For A1689-zD1, the HST F160W image is shown in the background of cutouts. 
    {\em Cutouts in the upper panels:} Gray and blue contours show moment-0 images of \Ciium\ and \Oium\ emission lines, respectively. 
    {\em Cutouts in the lower panels:} Gray and orange contours show moment-0 images of \Ciium\ and \Niium\ emission lines, respectively. 
    For all cutouts, solid contours show $3, 5, 7, 9\,\sigma$ for \Cii\ and $2, 3, 4, 5\,\sigma$ for \Oi\ and \Nii\ data. 
    Dashed contours show $-2,-3,-4,-5\,\sigma$ signals (if exist). 
    {\em Spectra}: Gray histograms show \Ciium\ lines. Blue and orange hatched histogram show \Oium\ and \Niium\ lines. Flux densities of the spectra are scaled by arbitrary factors. While \Oi\ emission lines of all the targets are detected with the significance of $>4\,\sigma$, \Nii\ emission lines are non-detected except for the weak signal from A1689-zD1.}
    \label{fig:cutouts}
\end{figure*}

\renewcommand{\arraystretch}{1.2}
\begin{table*}
\begin{threeparttable}
    \caption{Summary of measurements of our samples}
    \label{tab:ALMAmeasurements}
    \centering
    \begin{tabular}{lcccc}
        \hline
           Name  &  REBELS-38 & A1689-zD1 & REBELS-25 & REBELS-18\\
        \hline
        Magnification factor & -- & $4.14 \pm 0.36 ^{\ast}$ & -- & -- \\
        \hline
           $z_{\rm [OIII]88}$ & -- & $7.132^{a}$ & $7.3052 \pm 0.0004^{i}$ & --\\
           FWHM$_{\rm [OIII]88}$ (${\rm km/s}$) & -- & $182\pm31^{a}$ & \red{$203\pm38^{i}$} & --\\
           f$_{\rm [OIII]88}$ (${\rm Jy\,km/s}$) & -- & $53.3 \pm 3.5^{b}$ & \red{$896^{+234,i}_{-207}$} & -- \\
           $L_{\rm [OIII]88}$ ($\times 10^8\,{\rm L_{\odot}}$) & -- & \red{$124\pm8^{b}$} & \red{$20.0^{+5.1,i}_{-4.6}$} & --\\
           $S_{\rm 88}$ (${\rm \mu Jy}$) & $ 334\pm 63^{e}$ & $ 1757\pm 279^{f}$ & $ 486\pm 122^{e}$ & -- \\
           \hline
           $z_{\rm [CII]158}$ &  $6.5770\pm0.0001$ & $7.132^{a}$ & $7.3065\pm0.0002$ & $7.6750\pm0.0001$\\
           FWHM$_{\rm [CII]158}$ (${\rm km/s}$) & $     204\pm13$ & $323\pm8^{c}$ & $322\pm12$ & $209\pm11$\\
           f$_{\rm [CII]158}$ (${\rm Jy\,km/s}$) & $1.58\pm0.12$ & $3.56\pm0.07^{b}$ & $1.28\pm0.04$ & $0.81\pm0.05$\\
           $L_{\rm [CII]158}$ ($\times 10^8\,{\rm L_{\odot}}$) & $16.8\pm1.6$ & $45.0\pm1.0^{b}$ & $15.9\pm1.0$ & $10.8\pm0.9$\\
           $S_{\rm 158}$ (${\rm \mu Jy}$) & $ 163\pm 23^{d}$ & $ 558\pm 102^{g}$ & $ 260\pm 22^{d}$ & $ 53\pm 10^{d}$ \\
           \hline
            $T_{\rm d}$ (${\rm K}$) & $46^{+18}_{-11}$$^{j}$ & $42^{+13}_{-7}$$^{h}$ & $32^{+9}_{-7}$$^{i}$ & $39^{+12}_{-7}$$^{j}$ \\
            $\beta_{\rm d}$ & & $1.61^{+0.60}_{-0.75}$$^{e}$ & $2.5\pm0.4$$^{i}$ & \\
            $L_{\rm IR}$ ($\times 10^{11}\,{\rm L_{\odot}}$) & $8.0^{+4.4}_{-2.9}$ & $20.1\pm0.7^{a}$ & $5.0^{+2.9}_{-1.0}$$^{i}$ & $3.5^{+2.0}_{-1.3}$\\
        \hline\hline
        \multicolumn{5}{c}{This Work}\\
        \hline
            $z_{\rm [OI]145}$ & $6.5774\pm0.0003$ & $7.134\pm0.0006 $ & $7.308\pm0.001$ & $7.676\pm0.0006$\\
            FWHM$_{\rm [OI]145}$ (${\rm km/s}$) & $112\pm24$ & $278\pm51$ & $440\pm95$ & $165\pm38$\\
            f$_{\rm [OI]145}$ (${\rm Jy\,km/s}$) & $0.110 \pm 0.022$ & $0.556 \pm 0.137$ & $0.394 \pm 0.082$ & $0.107 \pm 0.028$\\
            $L_{\rm [OI]145}$ ($\times 10^8\,{\rm L_{\odot}}$) & $1.3 \pm 0.3$ & $7.3 \pm 1.8$ & $5.3\pm1.1$ & $1.6 \pm 0.4$\\
            $S_{\rm 145}$ (${\rm \mu Jy}$) & $220\pm28$ & $705\pm76$ & $270\pm41$ & $116\pm12$ \\
        \hline
            f$_{\rm [NII]205}$ (${\rm Jy\,km/s}$) & $<0.045$ ($3\,\sigma$) & $<0.059$ ($3\,\sigma$) & $< 0.076$ ($3\,\sigma$) & --  \\
            $L_{\rm [NII]205}$ ($\times 10^8\,{\rm L_{\odot}}$) & $<0.37$ ($3\,\sigma$) & $< 0.55$ ($3\,\sigma$) & $<0.73$ ($3\,\sigma$) & --\\
            $S_{\rm 205}$ (${\rm \mu Jy}$) & $131\pm39$ & $269\pm15$ & $94\pm12$ & --\\
        \hline
            $f_{\rm [CII]}^{\rm PDR}$ & \red{$>0.92$} & \red{$>0.96$} & \red{$>0.83$} & --  \\
        \hline
    \end{tabular}
    \begin{tablenotes}
	\item  a. \citet{Akins2022}, b. \citet{Killi2023}, c. \citet{Wong2022}, d. \citet{Inami2022}, e. \citet{Algera2023}, f. \citet{Inoue2020}, g. \citet{Watson2015}. h. \citet{Bakx2021}, i. \citet{Algera2024}. j. \citet{Sommovigo2022}.
    \item $^{\ast}$ \red{See Appendix \ref{sec:lens} for the discussion of the magnification factor.} All the measurements from A1689-zD1 are not corrected for the lensing magnification.
    \end{tablenotes}
    \end{threeparttable}
\end{table*}

\subsection{Emission Line Detections}

For the \Oium\ emission line data, we find clear detections of emission lines at the expected frequencies from all targets.
\red{For the detected \Oi\ lines, we created  moment-0 images by determining the appropriate frequency widths that capture \Oium\ emission using an iterative method.
First, we performed 1D Gaussian fits on the extracted spectra and re-integrated the data cube using frequency ranges that encompass the $2\,\sigma$ line widths.
Next, we re-extracted spectra from regions in the moment-0 images where pixel values exceeded $2\,\sigma$, further refining the emission line spectra.
We then repeated the 1D Gaussian fitting to update the frequency width and generate new moment-0 images.
This process was repeated until the selected channels for moment-0 image creation converged.}

For the \Niium\ emission line data of REBELS-38 and REBELS-25, we did not identify any emission lines from the extracted spectra.
To further examine the \Nii\ lines, we created moment-0 images by using velocity widths fixed to those of \Ciium\ emission lines ($1\times$ FWHMs and $2\times$ FWHMs).
We only found signals $\lesssim3\,\sigma$ at the expected source position.
We thus concluded that all the \Niium\ emission lines were non-detected.

For \Niium\ emission line of A1689-zD1, we found a weak \red{$\lesssim3\sigma$} signal in its moment-0 map integrated over the same velocity FWHM as the \Ciium\ emission line \citep{Wong2022}. However, we treated the \Niium\ of A1689-zD1 as an upper limit based on the following reasons.
In the spectra (bottom panels in figure \ref{fig:cutouts}), the positive signal has two peaks that are slightly off from the expected frequency from the \Cii\ emission detection.
As clumpy structures of A1689-zD1 was reported \citet{Wong2022} and the spectra have a non-Gaussian shape, the positive signal might arise from some of the clumps distributed in a line of sight in the A1689-zD1 and the \Nii\ emitting region do not overlap with that of \Cii\ emission line.
As our current observations of \Niium\ and \Oium\ does not have enough spatial resolution to resolve the clumps and as the detection signals are marginal, we treat the \Niium\ signal of A1689-zD1 as an upper limit. However, even if we treat the \Niium\ as detection, our results do not change.

We measured line fluxes (for the detected lines) and flux upper limits (for non-detections) using the moment-0 images. For the detected \Oium\ emission lines, we performed 2-D Gaussian fittings using CASA task \path{imfit}.
For non-detected \Niium\ emission lines, we measured $3\,\sigma$ upper limit fluxes by calculating the $3\times$ pixel-by-pixel RMSs of moment-0 maps integrated over the FWHMs of \Cii\ emission lines.
All emission line images are shown in figure \ref{fig:cutouts} and our ALMA measurements are listed in table \ref{tab:ALMAmeasurements}.

\subsection{Dust Continuum Detection}
Dust continua were all clearly detected from all targets with $>4.5\,\sigma$ significance at the source positions.
We measured continuum fluxes by 2-D Gaussian fittings using the CASA task \path{imfit}.
Measured continuum fluxes are listed in table \ref{tab:ALMAmeasurements}.
Appendix \ref{sec:dust} shows dust continuum images.

\begin{figure*}
    \centering
    \includegraphics[width=1.01\textwidth]{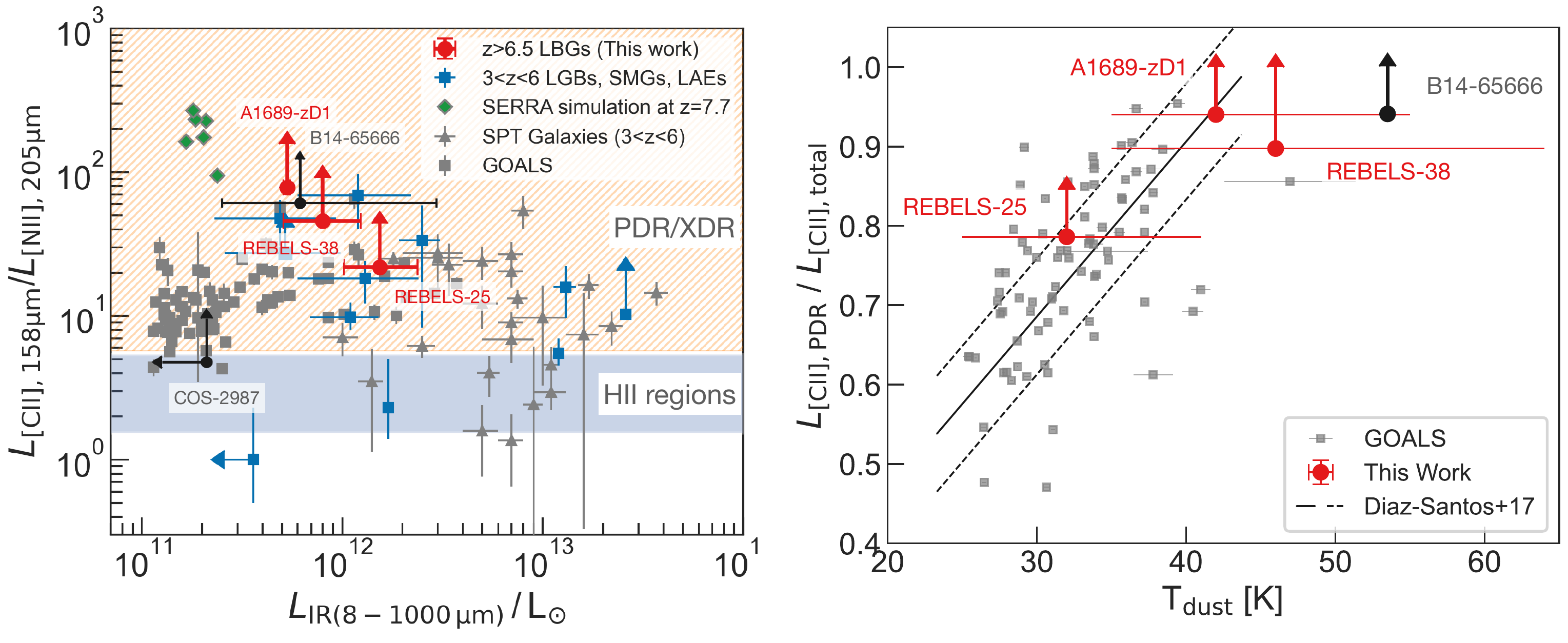}
    \caption{{\em Left Panel}: \Ciium\ over \Niium\ emission line luminosity ratio as a function of IR luminosity of local and high-redshift galaxies. Local galaxies are obtained from the GOALS sample \citep{Diaz-Santos2017}, while high-redshift galaxies at $z\sim3-6$ were studied in \citet{Decarli2014,Pavesi2019,Cunningham2020,Pensabene2021,Schreiber2021,Witstok2022} and in this work.
    Blue band represents \Ciium\ over \Niium\ emission line ratios expected from \Hii\ regions with a wide parameter ranges and the orange hatched band represents the line ratio from PDR or X-ray dominated region \citep{Witstok2022}.
    The $3\,\sigma$ lower limits of the $z>6.5$ galaxies (red points with arrows) are located well above most of the local and high-redshift galaxy distribution.
    The measured high ratios of \Ciium\ over \Niium\ emission line luminosity suggest that the \Cii\ emission lines from ISM of $z>6.5$ galaxies are dominated by those from PDRs.
    {\em Right Panel}: $f^{\rm PDR}_{\rm [CII]158}$ as a function of dust temperatures ($T_{\rm d}$). Gray squares are $z\sim0$ galaxies from GOALS \citep{Diaz-Santos2017}. Black solid line shows correlation between $f^{\rm PDR}_{\rm [CII]158}$ and $T_{\rm d}$ found in \citet{Diaz-Santos2017}.
    The $3\,\sigma$ lower limits found in this studies are consistent with the $z\sim0$ results, which reaches temperatures of up to $T_{\rm d}\sim40\,{\rm K}$.
    }
    \label{fig:C2N2ratio}
\end{figure*}

\section{Results} \label{sec:results}

Here we present the results of our observations, with particular focus on the inferred FIR line luminosities and the physics implied by their intrinsic ratios.

\subsection{IR Luminosity and FIR SED}
The FIR spectral energy distribution (SED) and IR luminosities\footnote{Throughout this work, $L_{\rm IR}$ are calculated using the rest-frame wavelength range of $8-1000\,{\rm \mu m}$.} of A1689-zD1 and REBELS-25, REBELS-38 have already been studied in detail by \citet{Bakx2021} and \citet{Algera2023}.
We adopt dust temperatures ($T_{\rm d}$), emissivity indexes ($\beta_{\rm d}$), and IR luminosities ($L_{\rm IR}$) measured in these previous studies.

REBELS-18 does not have direct FIR SED measurements as only low-spatial resolution $\lambda=158\,{\rm \mu m}$ continuum observations were obtained so far. With the new $\lambda=145$ and $205\,{\rm \mu m}$ continuum observations, the dust temperatures are challenging to constrain as these wavelengths only cover the Rayleigh–Jeans tail of the emission and the accurate constraints of dust temperatures require higher frequency  \citep[e.g., $\lambda_{\rm rest}\lesssim100\,\rm{\mu m}$;][]{Faisst2020} or higher spatial resolution observations \citep[e.g.,][]{Inoue2020,Fudamoto2023}.
In this study, thus we rely on the dust temperature and IR luminosity estimations for REBELS-18 inferred from the \Cii\ emission lines \citep{Sommovigo2022}.

\subsection{\Niium\ and \Ciium\ line luminosity ratio}

Using the non- or tentative-detections of the \Niium\ emission lines for our target galaxies, we calculated $3\,\sigma$ lower limit luminosity ratio $L_{\rm[CII]158}/L_{\rm [NII]205}$ ranging from $\gtrsim21$ to $\gtrsim87$ (left panel of figure \ref{fig:C2N2ratio}).
The observed line ratios of these EoR star-forming galaxies are generally much higher than the average distributions of local IR luminous galaxies observed by GOALS \citep{Diaz-Santos2017} and high-redshift SMGs at $z\sim4-6$ \citep{Cunningham2020}. 
Also, these high \Niium/\Ciium\ line ratios are consistent with the ${\rm SFR}>30\,{\rm M_{\odot}\,yr^{-1}}$ sample of simulated galaxies at $z=7.7$ from the SERRA simulation \citep{pallottini:2022}.
The $3\,\sigma$ lower limits of A1689-zD1 and REBLES-38 show the highest line luminosity ratios, which strongly indicate that the \Ciium\ emission lines are dominated by those from the PDR and not from HII regions (see \S 5.1).

\begin{figure*}
    \centering
    \includegraphics[width=0.99\textwidth]{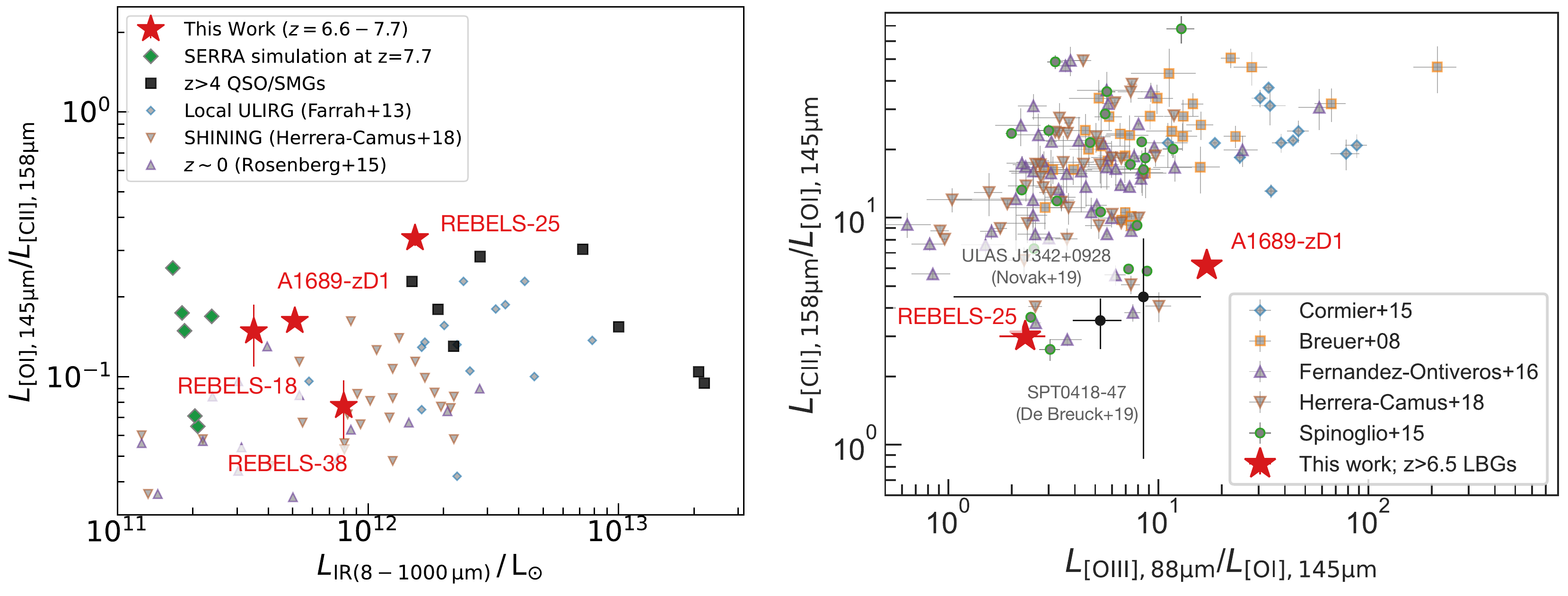}
    \caption{
    {\em Left Panel}: \Oium\ over \Ciium\ emission line luminosity versus IR luminosity for local and high-redshift galaxies and AGNs.
    Blue squares are from \citet{DeBreuck2019,Yang2019,Li2020,Lee2021,Meyer2022}, simulated results for $z\sim6$ galaxies are from \citep{Lupi2020}, and $z\sim0$ galaxies  are from \citet{Farrah2013,Spinoglio2015,Herrera-Camus2018,Rosenberg2015}. This work (red large stars) expanded \Oi\ emission line observations of high-redshift galaxies toward the lower IR luminosity range. The results show that \Oi/\Cii\ ratio of high-redshift galaxies have similar values to those of local galaxies, but tend to distribute at around the highest edge of the distribution.
    {\em Right Panel}: \Ciium\ over \Oium\ line luminosity  versus \Oiiium\ over \Oium\ line luminosity  for local- and high-redshift galaxies and AGNs. Seyferts \citep{Spinoglio2015}, local AGNs \citep{Fernandez-Ontiveros2016}, SHINING Survey \citep{Herrera-Camus2018}, local dwarf galaxies \citep{Cormier2015}, local galaxies and Seyferts \citep{Brauher2008}. High-redshift galaxy and AGNs are from \citet{Novak2019} and \citet{DeBreuck2019}.
    }
    \label{fig:O1C2-LIR}
\end{figure*}

\subsection{\Oium, \Ciium, and \Oiiium\ line luminosity ratio}

From the observed \Oium\ emission line luminosity, we measure the \Oium\ to \Ciium\ emission line ratios to be in a range between $0.08\pm0.02$ and $0.33\pm0.07$ with a median of $0.16$ (figure \ref{fig:O1C2-LIR}).
The range of the \Oi/\Cii\ ratio is similar to that of $z>4$ QSOs/SMGs.
Most of our observation results are distributed at around the higher edge of local galaxy distribution at a fixed IR luminosity range.
The high \Oi/\Cii\ line ratio indicate high density of neutral gas regions.
\red{High neutral gas density in high-redshift galaxies have been reported in previous observational studies \citep[e.g., ${\rm log(}n{\rm /cm^{-3})}>4$; ][]{Bothwell2017}, although the emission line used and the neutral gas regions investigated would differ from those in our  study.}
We will discuss the implications of these \Oium\ observations in detail in \S\ref{sec:ISM}

We also investigated \Ciium/\Oium\ luminosity ratio versus \Oiiium/\Oium\ luminosity ratio (right panel of Figure \ref{fig:O1C2-LIR}) for REBELS-25 and A1689-zD1 as these galaxies have \Oiiium\ emission line observations reported in previous studies.
\red{In this diagram REBELS-25 is located at the lower left corner i.e., the lowest ranges of \Oiiium/\Oium\ ratio as well as \Ciium/\Oium\ ratio. In contrast, A1689-zD1 shows a higher \Oiiium/\Oium\ ratio compared to most local galaxies/AGNs, similar to the local dwarf galaxies studied by \citet{Cormier2015}.
The \Oiiium/\Oium\ line ratio is correlated with the volume filling factor of ionized oxygen, which in turn is related to the ionization parameter ($U_{\rm ion}$) \citep{DeBreuck2019}.
}
\red{
Previous studies have shown that the high [OIII]/[CII] line ratio in high-redshift galaxies is due to a high $U_{\rm ion}$ and/or lower PDR covering factors \citep[e.g.,][]{Hashimoto2019, Harikane2020, Witstok2022}. REBELS-25 and A1689-zD1 have [OIII]/[CII] ratios of 1.3 \citep{Algera2023} and 2.1 \citep{Wong2022}, respectively. The [OIII]/[CII] ratio of REBELS-25 is among the lowest values observed for $z > 6$ star-forming galaxies.
The low \Oiiium/\Oium\ line ratio of REBELS-25, suggesting low $U_{\rm ion}$, is consistent with its relatively low \Oiiium/\Ciium\ line ratio. 
In contrast, the higher \Oiiium/\Oium\ line ratio of A1689-zD1 suggests a higher $U_{\rm ion}$, which is generally consistent with its higher \Oiiium/\Ciium\ ratio. 
Although sample size is currently too small,  these line ratios suggest the potential for using multiple FIR emission lines to diagnose gas properties.
}

\subsection{\red{Origin of \Oi\ and \Cii\ emission lines of REBELS-38}}
\red{
Most of the target galaxies have \Oium\ emission lines that spatially coincide and have profiles consistent with those of [CII] emission lines.
However, \Oi\ emission line of REBELS-38 has significantly ($>2\,\sigma$) narrower emission line profile compared with \Cii\ emission line (i.e., ${\rm FWHM_{[OI]}=112\pm24}$ vs ${\rm FWHM_{[CII]}}=204\pm13\,{\rm km/s}$). The difference of the emission line profile suggests that the \Oi\ and \Cii\ of REBELS-38 are originated from different ISM components. In such case, analysis and discussions using \Oi/\Cii\ of REBELS-38 may breaks down.
To study this problem in detail, higher resolution observations of both \Oi\ and \Cii\ emission lines are necessary. Thus, we only remark on this potential issue for REBELS-38.
}

\section{Discussion} \label{sec:discussion}

\subsection{Origins of \Ciium\ Emission Lines}
\label{sec:C2origin}

Following previous studies \citep[e.g.,][]{Decarli2023}, we describe the fraction of \Ciium\ emission originated from PDR using \Cii\ and \Nii\ luminosity ratio as: 
\begin{equation}
    f^{\rm PDR}_{\rm [C\sc{II}]} = 1 - R_{\rm ion}\,(\frac{L_{\rm{[N\sc{II}]205\,\mu m}}}{L_{\rm{[C\sc{II}]158\,\mu m}}})\\
\end{equation}
where $R_{\rm ion}$ is the expected \Cii\ over \Nii\ emission line luminosity ratio when \Cii\ is originating entirely from \Hii\ regions.
Due to the similar emission properties of \Ciium\ and \Niium\ (e.g., their critical density and excitation temperature), $R_{\rm ion}$ was found to have a tight range between $2.5$ to $3$ assuming an electron density of the  range $n_{\rm e} = 10$ - $100\,{\rm cm^{-3}}$ \citep{Croxall2017}. Using wider ranges of parameters (metallicity, ionization field strength, gas density), \citet{Witstok2022} found a wider range of  $R_{\rm ion}=1.5$ - $5.7$ \red{by incorporating various metallicities, ionization parameters, hydrogen densities, and elemental abundances} (\red{blue band in} Figure \ref{fig:C2N2ratio}).
In this study, we use $R_{\rm ion}=3.6$ as our fiducial value by taking an average of $R_{\rm ion}$ found in the analysis of \citet{Witstok2022}. Later, we will discuss how the results change by using a higher value of $R_{\rm ion}=5.7$.

With the measured luminosity ratio between \Ciium\ and \Niium\ of the target galaxies, we find $3\,\sigma$ lower limits of $f_{\rm [CII]}^{\rm PDR} > 0.92$, $>0.96$, and $>0.83$ using $R_{\rm ion}=3.6$ for REBELS-38, A1689-zD1, and REBELS-25, respectively.
We also estimated the PDR fraction of \Cii\ of B14-65666 at $z=7.15$ where \Niium~line luminosity is estimated from \Nii$122\,{\rm \mu m}$ line \citep{Hashimoto2019,Sugahara2021}.
With the estimated \Niium\ line luminosity assuming $n_{\rm e}=200\,{\rm cm^{-3}}$, we find $3\,\sigma$ lower limit of $f_{\rm [CII]}^{\rm PDR}>0.94$  for $R_{\rm ion}=3.6$, respectively. These $f_{\rm [CII]}^{\rm PDR}$ are thus similarly high as our \Niium\ observed samples.
Even using the large value of $R_{\rm ion}=5.7$, we find $f_{\rm [CII]}^{\rm PDR} > 0.87$, $>0.93$, $>0.74$, and $>0.91$ for REBELS-38, A1689-zD1, REBELS-25, and B14-65666, respectively.

These inferred PDR fractions of \Cii\ are high $f_{\rm [CII]}^{\rm PDR}$  compared with those of lower-redshift star-forming galaxies in the similar IR luminosity range  and higher 
 than those explained by ionized gas dominated emissions (left panel of figure \ref{fig:C2N2ratio}). 

\begin{figure*}
    \centering
    \includegraphics[width=1.0\textwidth]{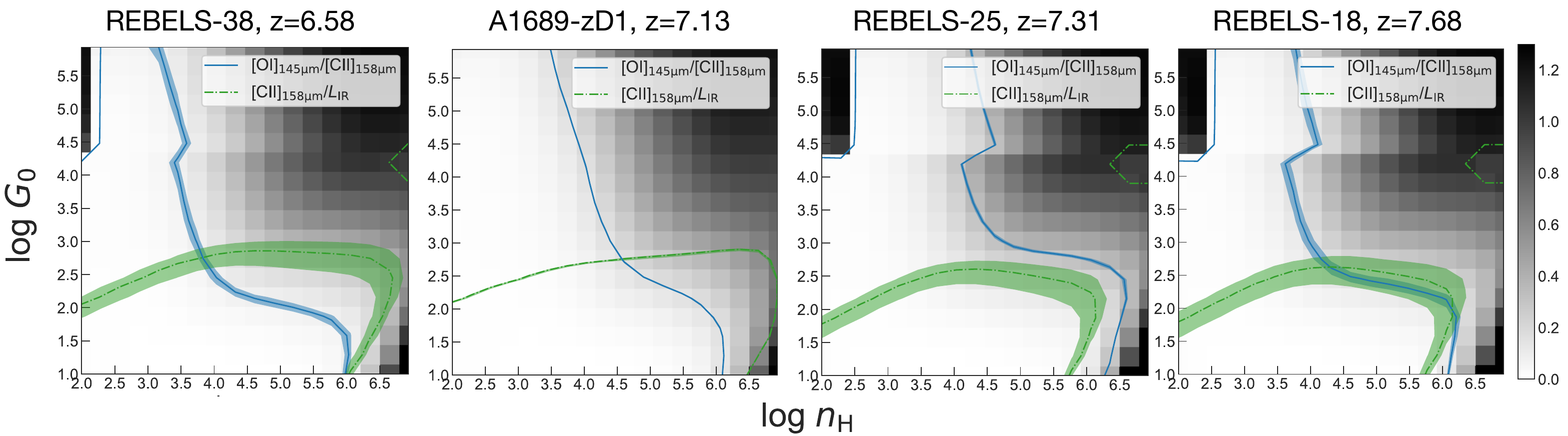}
    \caption{Grid of hydrogen gas density ($n_{\rm H}$ [$\rm{cm^{-3}}$]) versus interstellar radiation field ($G_0$) calculated using \path{CLOUDY} \citep{Ferland2017} assuming gas phase metallicity recently estimated for each galaxies from JWST observations and total gas column density of ${\rm log}(N_{\rm H}\,{\rm cm^{-2}})=23$.
    Background images show \Oium/\Ciium\ intensity ratio calculated in each grids. 
    Observed luminosity ratio for each galaxies are plotted by solid and dashed lines and $1\sigma$ uncertainties are shown by bands.
    While the estimated FUV field strengths are similar for these galaxies with $G_0\sim10^{2.5} - 10^{3.0}$ and PDR gas densities are high with $n_{\rm PDR}\sim10^{3.5}$ - $10^6\,{\rm cm^{-3}}$.}
    \label{fig:pdrt}
\end{figure*}

Previous studies of local IR luminous galaxies find that the dust temperature and the $f_{\rm [CII]}^{\rm PDR}$ correlate well \citep[e.g.,][]{Malhotra2001,Diaz-Santos2017}.
Our target galaxies are consistent with the correlation found in \citet{Diaz-Santos2017} except for the fact that most of the high-redshift and high-$T_{\rm d}$ galaxies studied here have saturated $f_{\rm [CII]}^{\rm PDR}$ (right panel of figure \ref{fig:C2N2ratio}).
\citet{Diaz-Santos2017} discussed that the correlation is due to the increase of the gas density in the star-forming region, which result in the increase of dust temperatures, in turn, which enhance the collisional de-excitation or the thermalization of \Cii\ and makes \Ciium\ emission faint especially in ionized region where the critical density of \Ciium\ emission is much lower than that in the neutral gas\footnote{The critical density of \Ciium\ emission is $\sim50\,{\rm cm^{-3}}$ and $\sim3\times10^{3}\,{\rm cm^{-3}}$ when the collision pair is the electron and the hydrogen atom/molecule, respectively \citep[e.g.,][]{Stacey2010,Carilli2013}.}.
As a result, high density and high temperature environment of ionized gas would make their contribution to the total \Ciium\ luminosity low.

On the other hand, recent studies find an increase in dust temperature as a function of redshift for normal star-forming galaxies \citep[e.g.,][]{Schreiber2018,Faisst2020,Sommovigo2022}. These studies find that the typical dust temperature of star-forming galaxies at $z\sim6-7$ to be $T_{\rm d}\sim40-60\,\rm{K}$.
If the correlation of $f_{\rm [CII]}^{\rm PDR}$ and $T_{\rm d}$ exist for the $z\gtrsim6$ galaxies and the dust temperature of star-forming galaxies systematically increase as a function of redshift, the high-redshift star-forming galaxies should have very high or saturated $f_{\rm [CII]}^{\rm PDR}$ by combining these two correlations.
Our finding of the high $f_{\rm [CII]}^{\rm PDR}$ from our target galaxies are consistent with the scenario in which high-redshift galaxies have the high $f_{\rm [CII]}^{\rm PDR}$ due to high gas densities and/or high dust temperatures.

\red{These features show that \Ciium\ emission lines from star-forming galaxies in the epoch of reionization are more dominated by PDR regions compared with lower redshift galaxies, as also suggested by \citet{Heintz22,Heintz23c}. Thus, we conclude that the origin of \Ciium\ emission lines of these star-forming galaxies studied here are neutral gas. This is in line with simulations \citep[e.g.,][]{Pallottini2017} and theoretical predictions \citep[e.g.,][]{Vallini2017,Ferrara2019}. These results indicate that \Ciium\ emission is an excellent tracer of neutral ISM.}

We hereafter treat all the \Cii\ emissions from our targets are all come from their neutral gases, and no corrections are applied to study when combining with \Oium\ emission line luminosity in the following sections.

\subsection{ISM properties}
\label{sec:ISM}

\subsubsection{CLOUDY settings and grids}
To analyze the neutral gas conditions of our sample, we used \path{CLOUDY} \citep[version 17.03;][]{Ferland2017} with the observed ratios of fine structure lines.
In this analysis, we focus on modeling emission only from neutral gas (i.e., PDR and molecular gas) and we treat \Ciium\ emission lines fully arising from neutral gas as we find in Section \ref{sec:C2origin}.
To make models of neutral gas emission using \path{CLOUDY}, we removed ionizing photons from impinging radiation fields\footnote{We allowed the ionizing radiation fully absorbed by a large column of gas ($N{\rm(H)}=10^{24}\,{\rm cm^{-2}}$) without any leakage using the \path{CLOUDY} input of \texttt{extinguish by 24 leakage=0}.}.

We fixed the impinging radiation to blackbody radiation with a temperature of $T=50000\,\rm{\rm K}$ by setting \texttt{blackbody 5e4 K} in the \path{CLOUDY} input, following previous studies \citep{Pensabene2021,Decarli2023}.
The FUV radiation field intensity was defined using Habing flux in the unit of $\rm{G_0}$ where $1\,{\rm G_0}$ corresponds to $1.6\times10^{-3}\,{\rm erg\,s^{-1}\,cm^{-2}}$ over $6<h\nu<13.6\,{\rm eV}$ \citep{Habing1968}. We then made a grid of $G_0$ ranging $1<{\rm log}\,G_0<6$ and the ${\rm log}\, G_0$ is spaced by $0.29\,{\rm dex}$. 
Cosmic rays are added as a background, which works for the ionization balance in deep neutral-molecular gas regions. We used the \path{CLOUDY}'s default setting by using the command: \texttt{cosmic rays, background}.
We used the plane-parallel geometry with a constant hydrogen density by setting \texttt{constant density}.
The elemental and the dust grain abundance is adopted using \path{CLOUDY}'s default set for ISM by setting \texttt{abundance ISM}.
The polycyclic aromatic hydrocarbons (PAHs) abundance is then set by using \texttt{grains PAH}.
The grain and gas metallicity is scaled to values based on the oxygen abundance mesurements by recent JWST observations; $12+{\rm log(O/H)}= 8.45$, $8.05$, $8.63$, and \red{$8.50$} for REBELS-38, A1689-zD1, REBELS-25, and REBELS-18, respectively  (\citealt{Rowland2025} and Carmen Blanco-Prieto; private communications).
\red{Finally, we set a stopping criteria of reaching a gas column density of $\rm{log(}N_{\rm H}/{\rm cm^{-2}})=23$.}
We executed all the \path{CLOUDY} run until it converges using \texttt{iterate to converge}.
The neutral gas density ranges $2<{\rm log}\,n_{\rm H}/{\rm cm^3}<7$ spaced by $0.29\,{\rm dex}$. 

With the above \path{CLOUDY} setups, we made a grid of emission line models by varying hydrogen density $n_{\rm H}\,{\rm cm^{-3}}$ and radiation field strength. We then calculated emission ratios of \Cii/$L_{\rm IR}$, and \Cii/\Oi\ for all $n_{\rm H}\,{\rm cm^{-3}}$ and FUV radiation field strength\footnote{We publish input files and output grids of our cloudy run:\url{https://github.com/yfudamoto/Fudamoto_OI_cloudy.git}}.

\subsubsection{Gas densities and FUV field strengths of star-forming galaxies in the EoR}

\begin{figure}
    \centering
    \includegraphics[width=0.9\columnwidth]{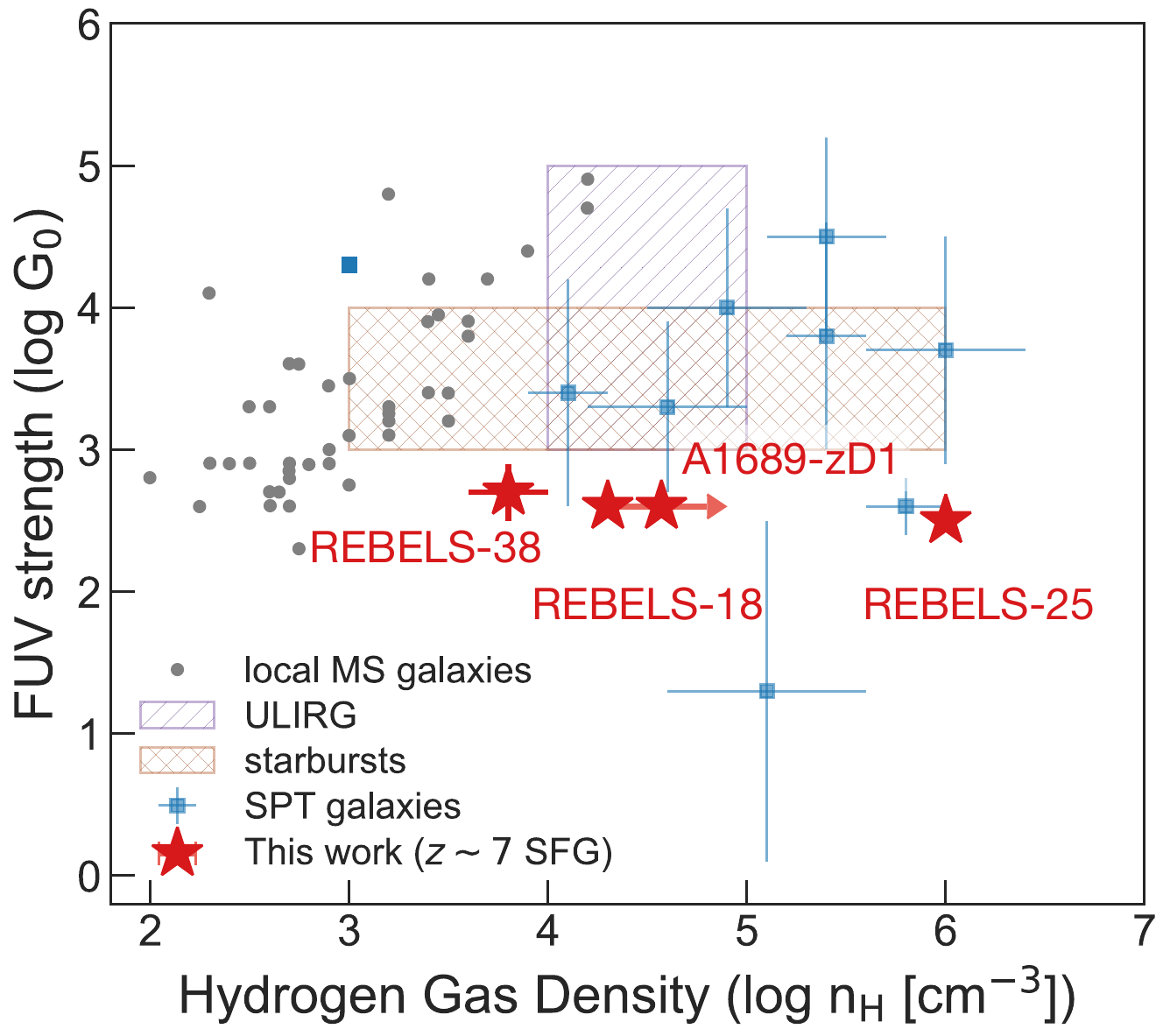}
    \caption{
    Hydrogen gas density ($n_{\rm H}$) versus FUV field strength ($G_0$) of local and high-redshift galaxies.
    Gray dots show local main-sequence galaxies \citep{Malhotra2001}.
    $n_{\rm PDR}$ and $G_0$ range of local ULIRGs and starbursts are shown using purple and green hatched squares, respectively \citep{Stacey1991,Davies2003}.
    Blue squares are $z\sim4$ dust-obscured starbursts observed as part of the SPT survey \citep{Bothwell2017,DeBreuck2019}.
    Our target galaxies, \red{\Cii\ bright} star-forming galaxies in the EoR, have higher ($\times\sim10$ to $\times\sim1000$) gas density than local main-sequence galaxies, while UV field strengths are similar to local main-sequence galaxies and lower than those of starbursts.
    }
    \label{fig:nG0}
\end{figure}

Figure \ref{fig:pdrt} shows our results of the \path{CLOUDY} modelings with measured \Oium/\Ciium\ and \Cii/$L_{\rm IR}$ emission ratios.
All galaxies except for REBELS-25 have solutions of $G_0$ and $n_{\rm H}$ that explain the observed line ratios.
We find that radiation field strengths have similar values ranging around ${\rm log}\,G_0\sim2.5 - 3.0$.
This range of radiation field strength is a factor of a few to up to an order of magnitude lower than those found in high-redshift SMGs or quasars \citep[see figure \ref{fig:nG0}; ][]{Bothwell2017,DeBreuck2019,Novak2019,Pensabene2021,Meyer2022}.
On the other hand, the gas densities range ${\rm log}\,n_{\rm H}\,{\rm cm^{-3}}\sim 3.5$ to $\sim6$ that are $\times10$ to $\times1000$ higher than those of local main-sequence galaxies and that are similar to those of local, high-redshift starbursts, or SMGs , which is consistent with the finding of \citet{Markov2022}.
Overall, compared with high-redshift ($z\sim4-6$) SMGs, our target star-forming galaxies in the EoR have a similar range of gas densities and lower FUV radiation field strengths (figure \ref{fig:nG0}).

The lower $G_0$ and similar $n_{\rm H}$ compared with SMGs could be simply explained by the lower SFRs of our targets than those of SMGs and similar gas densities.
Assuming a spherical distribution of neutral gas at a fixed gas density, the mass of the neutral gas (and thus star-formation rates) scales as $\propto r^{3}$.
By defining $f$ as the mass ratio between starbursts and star-forming galaxies, output intensity from star-forming activities scale as $\propto f^{(1/3)}$ (see footnote\footnote{FUV intensity ($G_0$) and gas density ($n$) scales as $G_{\rm 0}\propto L/R^2$ and $n\propto M/R^3$, respectively, where $M$, $L$, and $R$ are the mass, the luminosity, and the size of the system. Assuming $L\propto M$ and assuming that $M$ and $L$ of starburst galaxies and star-forming galaxies are different by a factor of $f$ and the gas densities are same (i.e., $n_{\rm SB} = n_{\rm SFG}$, $L_{\rm SB}=f\,L_{\rm SFG}$, and $M_{\rm SB}=f\,M_{\rm SFG}$), we can find that $G_0$ scales as $f^{1/3}$.}).
Typically, dusty star-forming galaxies found by SPT survey have much larger SFRs (e.g., a few $\times 100\,{\rm M_{\odot}\,yr^{-1}}$) which is $\sim 5$ - $10\times$ larger than those of galaxies studied here. Thus, in this simple estimation, we expect $\sim5^{1/3}$ - $10^{1/3} \sim 2\times$ (or $\sim0.3\,{\rm dex}$) lower FUV intensity , if we fix a gas density, which is consistent with our results that show starbursts and star-forming galaxies in the EoR have a similar range of $n_{\rm H}$ and lower $G_0$.
Thus, our PDR analysis suggests the picture that \red{\Cii\ bright} star-forming galaxies in the EoR have similarly high gas densities with simply scaled-down star formation activities than those of SMGs.

\red{ The difficulty of explaining REBELS-25's emission line ratios is likely due to the uncertainty of the gas-phase metallicity of the neutral gas as we do find that clear solution for $G_0$ and $n_{\rm H}$ if we assume lower metallicity (e.g., $12+{\rm log(O/H)}=8.05$) rather than higher metallicity of  $12+{\rm log(O/H)}=8.63$.
Since our metallicity is estimated using ionized gas emission lines observed using JWST \citep{Rowland2025}, we assumed that same metallicity for ionized gas and neutral gas.
If this assumptions is not the case and the majority of neutral gas of REBELS-25 has lower metallicity (e.g., due to infalling pristine gas) than ionized gas, our \path{CLOUDY} model may explain the \Oi/\Cii\ and \Cii/$L_{\rm IR}$ with $G_0\sim10^{2.5}$ and $n_{\rm H}\sim10^{6}\,{\rm cm^{-3}}$.
While the $n_{\rm H}$ and $G_0$ for REBELS-25 are uncertain, the high \Oi/\Cii and high \Cii/$L_{\rm IR}$ indicate that REBELS-25 has low $G_0$ with high $n_{\rm H}$.
In the following we thus assume $G_0\sim10^{2.5}$ and $n_{\rm H}\sim10^{6.5}\,{\rm cm^{-3}}$ as the likely parameters for REBELS-25.}

\begin{table*}
    \centering
    \begin{tabular}{cccccc}
    \hline
      ID & FUV strength & $n_{\rm H}$  & $M_{\rm O, T_{\rm ex}=50\,{\rm K}}$ & $M_{\rm H\_[OI]145}$   \\
        & (${\rm G_0})$ & (${\rm cm^3}$) & ($\times 10^8\,{\rm M_{\odot}}$) & ($\times10^{9}\,{\rm M_{\odot}}$)\\
    \hline
       REBELS-38 & $10^{2.8\pm0.2}$ & $10^{3.8\pm0.2}$ & \red{$1.0\pm0.2$} & \red{$1.4\pm0.3$}    \\
       A1689-zD1 & $10^{2.76\pm0.02}$ & $10^{4.57\pm0.03}$ & \red{$1.4\pm0.4 \times (4.14/\mu)$}  & \red{$4.9\pm0.4 \times (4.14\mu$)}   \\
       REBELS-25 & $\sim10^{2.5}$ & $\sim10^{6.5}$ & \red{$4.2\pm0.9$} & \red{$3.9\pm0.8$} \\
       REBELS-18 & $\sim10^{2.6}$ & $>10^{4.3}$ & \red{$1.3\pm0.3$} & \red{$25\pm6$}  \\
    \hline
    \end{tabular}
    \caption{Neutral gas properties estimated in our observations by the \texttt{CLOUDY} models and by assuming excitation temperature of $150\,{\rm K}$ for \Oium\ emission line and measured oxygen abundances.}
    \label{tab:mass}
\end{table*}

\subsection{Hydrogen and Oxygen Mass}
\subsubsection{Oxygen mass}

In the optically thin limit\footnote{\Oium\ emission is optically thin in a wide range of parameter space \citep[e.g.,][]{Kaufman1999}.}, the oxygen atom masses of galaxies can be derived using \Oium\ luminosity and the following formula

\begin{equation}
    M_{\rm O}/{\rm M_{\odot}} = 6.19 \times 10^{-5}\,Q(T_{\rm ex})\,e^{329\,{\rm K}/T_{\rm ex}}\,L^{\prime}_{\rm [OI]145}
\end{equation}
 where $Q(T_{\rm ex})$ is the partition function and $Q(T_{\rm ex})=5+3\,e^{-329\,{\rm K}/T_{\rm ex}}+e^{-228\,{\rm K}/T_{\rm ex}}$ for the oxygen atom, $L^{\prime}_{\rm [OI]145}/{\rm K\,km\,s^{-1}\,pc^{2}}$ is the integrated \Oium brightness
temperature in units of , and $T_{\rm ex}/{\rm K}$ is the excitation temperature  \citep{Weiss2003,Weiss2005,Meyer2022}.

For the calculation, we assume \red{an excitation temperature of $T_{\rm ex} = 50\,{\rm K}$, which is derived by CO emission line observations \citep{Riechers2009}.}
We estimated neutral oxygen masses of our targets between \red{$1.0\times10^8\,{\rm M_{\odot}}$ and $4.2\times10^8\,{\rm M_{\odot}}$ (table \ref{tab:mass}).}

We note that if we assume a higher $T_{\rm ex}$ the estimated atomic oxygen masses become smaller: e.g., \red{$M_{\rm O, 150\,{\rm K}}=1.4\times10^{6}$ assuming  $T_{\rm ex}=150\,{\rm K}$ for REBELS-38.}
Thus, there would be \red{a large} systemic uncertainty to the estimated atomic oxygen masses from the assumed excitation temperature.

\subsubsection{Hydrogen masses of EoR star-forming galaxies}

With the estimated atomic oxygen masses, we estimate hydrogen masses of our target galaxies.
To do this, we used oxygen abundances recently estimated using optical \Oiii\ with  H$\beta$ lines from JWST observations (\citealt{Rowland2025}, Stefanon et al., 2024 in prep. and Blanco-Prieto et al., 2024 in prep).
For this purpose, we applied oxygen abundances from JWST observations that are assumed during the \texttt{CLOUDY} calculations\footnote{$12+{\rm log(O/H)}= 8.45$, $8.05$, $8.63$, and $8.53$ for REBELS-38, A1689-zD1, REBELS-25, and REBELS-18, respectively}.
By applying the oxygen abundance to the derived oxygen mass, we find $M_{\rm H}$ between \red{$1.4\pm0.3\times10^{9}\,{\rm M_{\odot}}$ and $2.5\pm0.6\times10^{10}\,{\rm M_{\odot}}$ (table \ref{tab:mass}).}
With the estimated hydrogen mass, we estimate gas mass fractions of \red{$M_{\rm H}/(M_{\ast} + M_{\rm H})\sim0.3-0.9$}, that are similar ranges recently derived from high-redshift galaxy studies using \Ciium\ lines \citep{2010ApJ...714L.162H,Dessauges2020,Heintz22,Aravena2023,Heintz2023}.

\citet{Wilson2023} reported a tight correlation between \Oium\ emission line luminosities and \Hi\ gas masses in $z\sim2$ - $6$ star-forming galaxies using 
gamma-ray bursts as a probe, and supported by local galaxy observations and hydro-dynamical simulations. Using the correlation, we find \Hi\ gas masses of $5.9\times10^{9}$, $1.6\times 4.14/\mu \times10^{10}$, $2.3\times10^{10}$, $6.9\times10^{9}\,{\rm M_{\odot}}$ for REBELS-38, A1689-zD1, REBELS-25, and REBELS-18, respectively.
These derived values are in good agreements with our estimated hydrogen masses derived from the oxygen abundances and oxygen masses \red{assuming $T_{\rm ex}=50\,{\rm K}$, while there is a relatively large scatter likely arising from the uncertainties of assumed $T_{\rm ex}$}.

We also compared the estimated hydrogen masses with molecular gas masses estimated using \Ciium\ emission lines.
To do this comparison, \red{we used the relation found in \citet{2010ApJ...714L.162H}, who employed constraints on PDR models to derive the atomic gas mass of a hyperluminous ($L_{\rm IR}\sim10^{13}\,{\rm L_{\odot}}$) starburst galaxy at $z = 1.3$. Using the equation (1) of \citet{2010ApJ...714L.162H} and applying estimated gas density and gas temperature of $50\,{\rm K}$, we find atomic gas masses are $7.0\times10^9$, $4.0\times10^9$, $5.0\times10^9$, and $3.7\times10^9$ for REBELS-38, A1689-zD1, REBELS-25, and REBELS-18, respectively. We find the results are in good agreement with our estimations, although there are uncertainties of the ionization correction factor and the true C abundance.} 

We also used correlation found in \citet{Zanella2018}. From the correlation, we found molecular gas masses of $M_{\rm H_2}=5.2\times10^{10}$, $3.4\times10^{10}\times(4.14/\mu)$, $5.0\times10^{10}$, and $3.3\times10^{10}\,{\rm M_{\odot}}$ for REBELS-38, A1689-zD1, REBELS-25, and REBELS-38, respectively. This molecular gas masses are generally consistent but \red{an order of magnitude} higher than hydrogen mass we derived using \Oium\ emission above. Given the relatively large scatter of the \Cii-to-$M_{\rm H_2}$ conversion, these two estimation might still be consistent. However, these excess might represent some of the unaccounted oxygen mass exist in molecular phase or variations of conversion factor, $\alpha_{\rm [CII]}$ in our sample. Further detailed analysis is, however, beyond the scope of this paper.

\section{Conclusion} \label{sec:conclusion}

In this paper, we presented results of multiple FIR emission line observations and analysis of four main-sequence star-forming galaxies in the epoch of reionization at $z>6.5$. The target galaxies were selected based on their previous detections of luminous \Ciium\ emission lines ($L_{\rm [CII]}>10^{9}\,{\rm L_{\odot}}$).
We performed follow-up observations of \Oium\ and \Niium\ emission lines of the target galaxies. As results of the follow-up ALMA observations and analysis, we find the following:

\vspace{0.15cm}
\noindent$\bullet$ From the \Niium\ emission line observations, we only find non-detections and a tentative detections, showing that \Niium\ emission lines are much fainter than \Ciium\ emission lines ($3\,\sigma$ upper limit of $L_{\rm{[CII]158}}/L_{\rm [NII]205} >21 $ to $>87$).
From these high $L_{\rm{[CII]158}}/L_{\rm [NII]205}$ ratio, we estimate the fraction of \Ciium\ emission arising from ionized gas and find that the dominant fraction of \Ciium\ emission lines arise from neutral gas with $f^{\rm PDR} > 0.74$ to $>0.96$ ($3\,\sigma$ lower limits). These results indicate that \Ciium\ emission is an excellent tracer of neutral ISM.

\vspace{0.15cm}
\noindent$\bullet$ We detect \Oium\ emission lines from all the target galaxies and find \Oium/\Ciium\ luminosity ratio of $0.08$ - $0.33$ (median $0.16$).
The observations marks the first detections of \Oium\ emission lines and these are the first emission lines obtained purely neutral medium from \red{\Cii\ bright} star-forming galaxies in the epoch of reionization at $z>6.5$.
These observations show that \Oium\ emission line observations would be feasible for larger samples of galaxies to constrain neutral gas properties of high-redshift galaxies.

\vspace{0.15cm}
\noindent$\bullet$ Using spectral synthesis calculations of \path{CLOUDY}, we model neutral gas conditions of the target galaxies.
By calculating FIR emissions in a grid of gas densities and FUV radiation field strengths, we find that the target galaxies have similarly moderate FUV radiation field strength of $G_0\sim10^{2.5}-10^{3.0}$.
We also find high PDR gas densities of $n_{\rm PDR}\sim10^4$ to $\sim10^{6}\,{\rm cm^{-3}}$ that spans a similar range with high-redshift ($z\sim4-6$) SMGs.
Overall, the neutral ISM of \red{\Cii\ bright} star-forming galaxies studied here is characterized by having high gas densities and moderately strong FUV radiation fields: a lower $G_0$ version of high-redshift SMGs.
These features suggest that these star-forming galaxies similar gas density to SMGs but smaller sizes.

\vspace{0.15cm}
\noindent$\bullet$ We estimate neutral atomic oxygen masses using observed \Oium\ emission luminosities \red{while the derived values have large uncertainties of unknown individual $T_{\rm ex}$}. 
Then, by combining oxygen abundances recently obtained in recent JWST spectroscopy, we derive the mass of neutral hydrogen that resides in the same region (atomic and molecular gas) as neutral atomic oxygen.
The derived hydrogen masses are generally consistent with those derived using \Ciium\ emission lines.
The estimated gas mass ratios ($M_{\rm H}/(M_{\ast}+M_{\rm H})$) range from \red{0.3 to 0.8}, which is consistent with those derived in other studies at $4<z<8$.
These observations and analyses show that directly converting oxygen masses to hydrogen masses using JWST-derived oxygen abundances \red{might be possible and could be} a direct method to derive hydrogen masses in neutral gas regions.

\vspace{0.15cm}
The rapid progress of JWST observations opened a new era for galaxy studies in the EoR by finding large numbers of galaxies in the heart of cosmic reionization and by enabling observations of their rest-frame optical wavelength.
To fully understand galaxy growth at high redshift, however, observations of neutral gas and cold ISM tracers are essential and such observations are only possible using mm/submm facilities such as ALMA.
Through the first detections of \Oium\ from \red{star-forming galaxies in the EoR that are more representative of the typical galaxy population than SMGs or QSOs}, our pilot observations demonstrate how multi-line observations in FIR wavelength enable our access to gas properties in neutral gas regions.
Future surveys at FIR wavelengths, particularly \Oi\ emission lines, in conjunction with JWST observations will allow us to analyze multiple phases of gas, stars, and dust directly in high-redshift galaxies.

\begin{acknowledgments}
We thank Carmen Blanco-Prieto, Javier Arvarez-Marquez, Luis Colina, and the team of JWST observation (GO-1840) for important discussions about the JWST observations of A1689-zD1.
This paper makes use of the following ALMA data: ADS/JAO.ALMA\#2022.1.00446.S ALMA is a partnership of ESO (representing its member states), NSF (USA) and NINS (Japan), together with NRC (Canada), MOST and ASIAA (Taiwan), and KASI (Republic of Korea), in cooperation with the Republic of Chile. The Joint ALMA Observatory is operated by ESO, AUI/NRAO and NAOJ.
YF, YS, and AKI acknowledge support from NAOJ ALMA Scientific Research Grant number 2020-16B.
This work was supported by JSPS KAKENHI Grant Numbers JP22K21349 and JP23K13149.
PD acknowledges support from the NWO grant 016.VIDI.189.162 (``ODIN") and from the European Commission's and University of Groningen's CO-FUND Rosalind Franklin program. HI and HSBA acknowledge support from the NAOJ ALMA Scientific Research Grant Code 2021-19A. HI acknowledges support from JSPS KAKENHI Grant Number JP19K23462. RAAB acknowledges support from an STFC Ernest Rutherford Fellowship [grant number ST/T003596/1].
S.F. akcnowledges the funding from NASA through the NASA Hubble Fellowship grant \#HST-HF2-51505.001-A awarded by the Space Telescope Science Institute, which is operated by the Association of Universities for Research in Astronomy, Incorporated, under NASA contract NAS5-26555.
MA is supported by FONDECYT grant number 1252054, and gratefully acknowledges support from ANID Basal Project FB210003 and ANID MILENIO NCN2024\_112

\end{acknowledgments}

%

\vspace{5mm}
\facilities{ALMA, VISTA, HST}


\software{astropy \citep{Astropy}, Cloudy \citep{Ferland2017}
          }



\appendix

\section{Gravitational Lens Model of Abell 1689}
\label{sec:lens}
\red{Within the selected targets, A1689-zD1 is strongly lensed by the foreground galaxy cluster the Abell 1689. The magnification factor of A1689-zD1 is presented in \citet{Watson2015} using the pioneering mass modeling of the Abell 1689 \citet{Broadhurst2005, Limousin2007}. Since then, new spectroscopy data are published and redshifts of several multiple images are updated \citep{Bina2016}. Thus, we re-analyze the mass model of Abell 1689 using the updated multiple image data and using an updated lens modeling code.}

\red{Using the public code \path{glafic} \citep{Oguri2010,Oguri2021}, we estimate the mass model of the Abell 1689 cluster using the public multiple image catalogs \citep{Coe2010,Ghosh2023} removing unsecure multiple images.
Some of the spectroscopic redshifts of the multiple images are obtained using the MUSE observation results \citep{Bina2016}. We used 130 multiple images of 40 source galaxies, 14 of which have spectroscopic redshifts. 
The cluster is modeled with the five Navarro-Frenk-White (NFW) components with a fast approximation \citep{Oguri2021}, cluster galaxies \citep{Coe2010}, and external shear, and $m=3$ and $4$ multipole perturbations \citep{Oguri2010}. We then run \path{glafic} software to find the minimum $\chi^2$ assuming positional errors of $0\farcs8$ for multiple image positions. The relatively large positional errors reflect the fact that Abell 1689 is one of the most massive clusters in the Universe with the highly complex mass distribution. The root-mean-square of differences between observed and model-predicted multiple image positions is $0\farcs64$ for the best-fitting model.}

\red{From the updated mass model, we find a magnification factor for A1689-zD1 to be $\mu=4.14 \pm 0.36 $, which is smaller than that previously reported \citep[e.g., $\mu=9.3$;][]{Broadhurst2005,Watson2015}. The error is calculated using MCMC calculation around the best-fit solutions. 
Although the magnification factor is relatively uncertain due to the complex structure of the Abell 1689 cluster, we applied a gravitational lensing magnification of $\mu=4.14$ as our fiducial value throughout this study.
The exact value of the magnification factor does not strongly affect our conclusion, as the main focus of this study is line ratios.}




\section{Dust continuum detections}
\label{sec:dust}

\begin{figure*}
    \centering
    \includegraphics[width=0.9\textwidth]{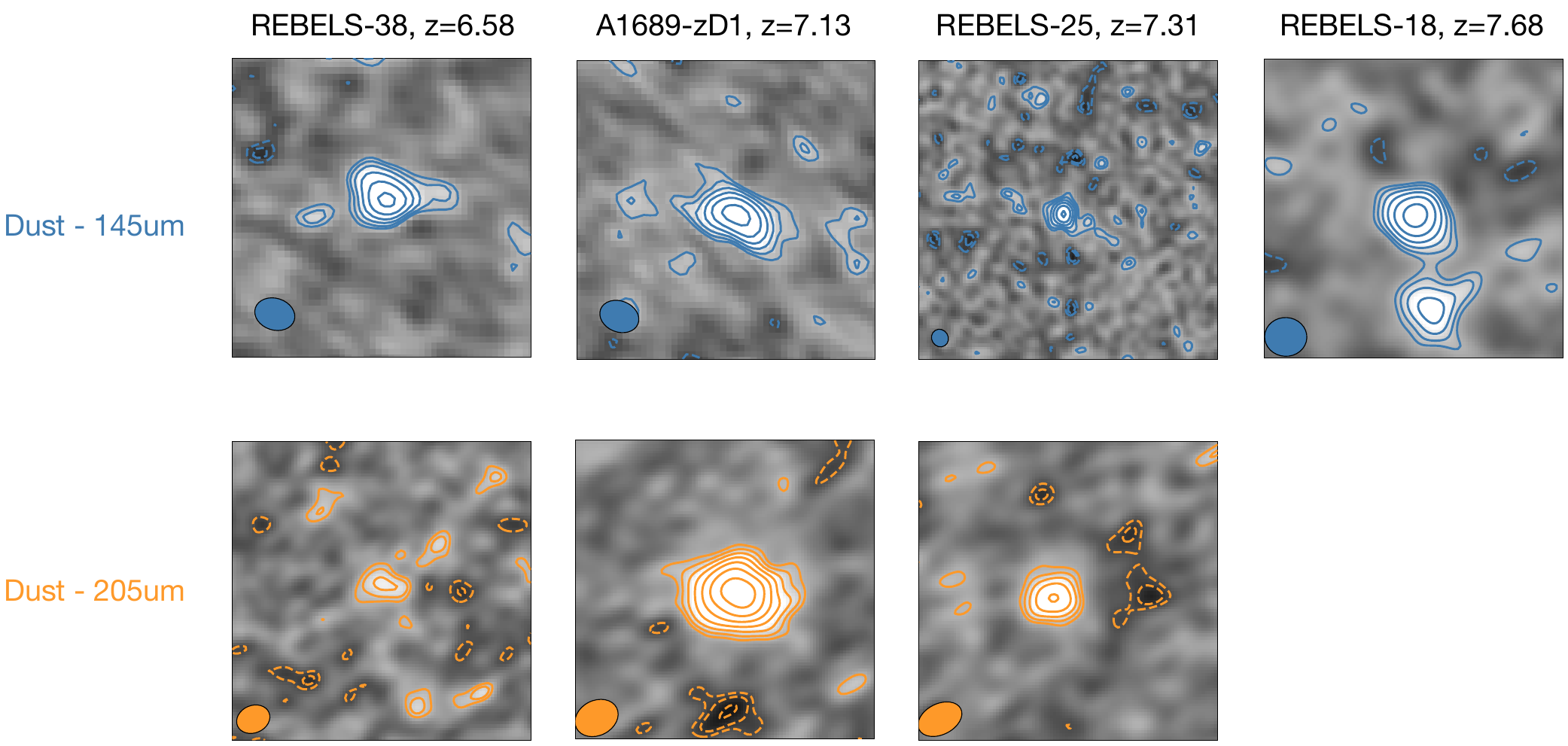}
    \caption{
    Dust continuum maps at $145\,{\rm \mu m}$ (upper panels) and $205\,{\rm \mu m}$
    (lower panels) for our target galaxies.
    All maps have $10\arc\times10\arc$ sizes.
    Solid contours show $2^{n}\,{\sigma}$ significance ($n=2,3,4,...$) and dashed contours show $-1\times2^{n}\,{\sigma}$ significance ($n=2,3,4,...$), if exist. Synthesized beam FWHMs are shown in the lower left corner of the plots.
    All continuum maps show significant ($>4.5\,\sigma$) detections of dust continua. Measured fluxes are listed in Table \ref{tab:ALMAmeasurements}. The off-center detection in the $145\,{\rm \mu m}$ dust image of REBELS-18 is a foreground galaxy with an estimated redshift of $z_{\rm ph}=3.1$ found in the COSMOS2020 catalog \citep{Weaver2023}.}
    \label{fig:continuum}
\end{figure*}

Here, we summarize images of dust continuum map of the $\lambda_{\rm rest}=145\,{\rm \mu m}$ and $\lambda_{\rm rest}=205\,{\rm \mu m}$ data of the target galaxies. All the dust continua are detected with $>8\,\sigma$ (Figure \ref{fig:continuum}). Measured fluxes are summarized in Table \ref{tab:ALMAmeasurements}.

\bibliography{base}{}
\bibliographystyle{aasjournal}



\end{document}